\documentclass[aps,prl,reprint,amsmath,amssymb,groupedaddress,superscriptaddress]{revtex4-1}
\usepackage{color}
\usepackage{enumitem}
\usepackage{graphicx}
\usepackage[normalem]{ulem}
\usepackage{comment}

\newif\ifHighlitedChanges
\def\ifHighlitedChanges{\iftrue}
\ifHighlitedChanges
  
  \def\STRIKE#1{{\color{red}\sout{#1}}}
\else
  
  \def\STRIKE#1{\relax}
\fi

\begin{document}

\bibliographystyle{apsrev}

\title{Wiedemann-Franz Law for Molecular Hopping Transport}
\author{Galen T. Craven}
\affiliation{Department of Chemistry, University of Pennsylvania, Philadelphia, PA  19104, USA} 
\author{Abraham Nitzan}
\affiliation{Department of Chemistry, University of Pennsylvania, Philadelphia, PA  19104, USA} 
\affiliation{School of Chemistry, Tel Aviv University, Tel Aviv 69978, Israel}

\begin{abstract}
The Wiedemann-Franz (WF) law is a fundamental result in solid-state physics that relates the thermal and electrical conductivity of 
a metal.
It is derived from the predominant origin of energy-conversion in metals: the motion of quasi-free charge-carrying particles. 
Here, an equivalent WF relationship is developed for molecular systems in which charge carriers are moving not as free particles but instead hop between redox sites.
We derive a concise analytical relationship between the electrical and thermal conductivity generated by electron hopping in molecular systems
and find that the 
linear temperature dependence of their ratio as expressed in the standard WF law is replaced by a linear dependence 
on the nuclear reorganization energy
associated with the electron hopping process.
The robustness of the molecular WF relation is confirmed by 
examining
the 
conductance properties of a paradigmatic molecular junction.
This result opens a new way to analyze conductivity in molecular systems, with possible applications advancing the design of molecular technologies that derive their function from electrical and/or thermal conductance.
\end{abstract}

 \maketitle

Electrical and thermal conductance 
are the 
principal
transport
mechanisms
giving rise to 
both 
the functionality 
and limitations
of the majority of devices
in the modern technological infrastructure.
Because of this importance,
understanding the interplay between these mechanisms is paramount in the development of next-generation
technologies. 
In metals, the relation between electrical conductivity $\sigma$ and the electronic contribution to the thermal conductivity $\kappa$ 
can be described 
using the Wiedemann-Franz (WF) law \cite{WFlaw1853}:
\begin{equation}
\label{eq:WFmetal}
\frac{\kappa}{\sigma} = \frac{\pi^2}{3} \left(\frac{k_\text{B}}{e} \right)^2 T = L T,
\end{equation}
where $L = \tfrac{\pi^2}{3} (k_\text{B} / e)^2$ is the Lorenz number,
$e$ is electron charge, $k_\text{B}$ is the Boltzmann constant, and $T$ is temperature.
The WF law forms the basis for analyzing the conductivity of myriad systems and materials 
in which the dominant energy transport mechanism is the nearly-free motion of charge-carriers \cite{Wakeham2011,Beneti2012,Crossno2016,Klockner2017,Cui2017,Lee2017science,Burkle2018}.
There is no equivalent law, however, for molecular systems where electron transport is often dominated by inelastic hopping between redox sites, and therefore
the theoretical picture of a possible relationship between the electric and thermal conductivities
of such systems
is currently lacking.

The analysis of electrical conductance in single-molecule devices is a well-established field 
that has primarily 
been motivated by the goal of designing and fabricating new nanoscale molecular electronics \cite{Ratner1974rectifier,Carroll2002,Nitzan2003electron,Ratner2013review,Xiang2016,Batista2017}. 
In contrast, the examination of thermal conductance that arises from electron transfer (ET) between molecules is a nascent research area.
A recent focus on electronic thermal conductance in molecular systems has been 
brought about 
by three particular advances: (a) newly developed experimental techniques that allow the measurement of temperature gradients over length scales commensurate with the distances involved in molecular ET processes \cite{Sadat2010,Menges2016,Mecklenburg2015},
(b) the development of a theory that describes electron-transfer-induced (ETI) heat transport, a phenomenon in which ET across a thermal gradient between molecules induces a heat current \cite{Segal2005prl,craven16c,matyushov16c,craven17a,craven17b,craven17e,craven18b,Kelly2019},
and (c) 
recent experiments in which
the thermal conductivity of a single molecule has been measured for the first time \cite{Reddy2019nature}.
These advances serve as impetus for analyzing the relations between different types of conductivity in molecules.

In this Letter, the ratio between ETI thermal conductivity and electrical conductivity is derived for 
purely molecular systems and then numerically validated in a molecular junction---a device
that is broadly applied in the development of nanotechnologies.
We term this ratio a \textit{molecular} Wiedemann-Franz (MWF) law, 
in analogy with the WF law for metals.
Experimental examination of this MWF relation could be performed using typical molecular junction setups \cite{Reddy2007,Tan2011,Lee2013,Kim2014,Venkataraman2015,Garner2018} by measuring the thermal conductance properties of a junction with suppressed phononic thermal conductivity and/or by taking two different measurements, first using conducting leads and then insulating leads, where the difference between them will be the ETI thermal conductivity \cite{Solomon2014,Li2017}.

To derive a MWF law \cite{note1}, consider a system consisting of identical molecular charge transfer sites at a density $\rho$ 
so that the distance between the center of the sites is $d = \rho^{-1/3}$. 
The electron transfer rate between identical molecular sites is given by the standard Arrhenius form
$k = k_0 e^{-E_\text{A} / k_\text{B} T}$ where $E_\text{A} = \lambda/ 4$ with $\lambda$ being the reorganization energy, an important physical parameter in condensed-phase ET reactions which parametrizes the electron-phonon coupling \cite{Marcus1956,Nitzan2006chemical}.
In the presence of an applied electric field $E$, 
the voltage difference between sites in the direction of the field is $ V = E d = E \rho^{-1/3}$.
Electrical conductivity $\sigma$ is defined from
\begin{equation}
\label{eq:voltgrad}
\text{J}_\text{el} = \sigma E = \sigma \frac{V}{d},
\end{equation}
where $\text{J}_\text{el}$ is the electrical current density. 
The electric current between a pair of identical hopping sites in the field direction is
\begin{equation}
\label{eq:eleccurrent}
\mathcal{J}_\text{el} = e \chi (1-\chi) k_0 \left( e^{-E_\text{A}/ k_\text{B} T} - e^{-\left(E_\text{A}+e V\right)/ k_\text{B} T}\right),
\end{equation}
where $\chi$ is the probability that a site is occupied by an electron, i.e., the fraction of occupied sites in the system, 
and the term $\chi (1-\chi)$ 
enforces the avoidance of multiple site occupancy.
The associated electrical conductivity is obtained as follows: Consider a volume element with unit surface area and thickness $d$. The surface density of sites in this volume element is $\rho_\text{s} = \rho d = d^{-2}$,
which is also a measure of the number of pairs across a surface in a picture where each site on one side of the dividing surface finds its nearest-neighbor on the other side of the surface. 
The electric current density in the field direction is
\begin{equation}
\begin{aligned}
\label{eq:elecdens}
\text{J}_\text{el} &= \mathcal{J}_\text{el} \rho_\text{s}\\
& =e \chi(1-\chi) \rho_\text{s}  k_0 \left( e^{-E_\text{A}/ k_\text{B} T} - e^{-\left(E_\text{A}+e V\right)/ k_\text{B} T}\right) \\
&\approx \frac{e^2 \chi(1-\chi) \rho  d^2 k  E}{k_\text{B} T},
\end{aligned}
\end{equation}
which implies that the electrical conductivity of the system is
\begin{equation}
\label{eq:eleccond}
\sigma = \frac{e^2 \chi (1-\chi) \rho  d^2  k}{k_\text{B} T}.
\end{equation} 
Next, consider the same system in the absence of an electric field but in the presence of a temperature gradient $\nabla T$. 
The thermal conductivity $\kappa$ is defined from 
\begin{equation}
\label{eq:Fourier}
\mathbf{J}_\mathcal{Q} = -\kappa \nabla T,
\end{equation}
where $\mathbf{J}_\mathcal{Q}$ is the heat current density.
In the direction of the temperature gradient, the heat current density is $\text{J}_{\mathcal{Q}} = -\kappa \nabla_{\! x} T$  
where $\nabla_{\! x} T = \Delta T / d$ with $\Delta T$ being the temperature difference between nearest sites.
We want to calculate the thermal conductivity in this system due to electron hopping between sites with different temperatures. 
To this end, consider two neighboring charge transfer sites $a$ and $b$ which have different local temperatures $T_a = T + \Delta T/2$ and $T_b = T - \Delta T/2$ but are otherwise identical and thus have the same electron occupation energy $E'$ and reorganization energy $\lambda$. 
In the absence of a driving electric field, the number of electron hops from site $a$ to site $b$ and from site $b$ to site $a$ per unit time is 
$\chi (1-\chi) k$.
Each such hop (in any direction) is associated with heat transfer $\mathcal{Q}$, the amount of which is given by \cite{craven16c}
\begin{equation}
\mathcal{Q}_{a,b} = \mathcal{Q}_{b,a} =  \lambda \left(\frac{ \displaystyle T_b - T_a   }{ \displaystyle T_a + T_b }\right) = -\frac{\lambda}{2} \frac{\Delta T}{T}.
\end{equation}
Hence, the heat current 
between sites is 
\begin{equation}
\label{eq:heatcurrent}
\mathcal{J}_\mathcal{Q} =\chi (1-\chi) k  \left(\mathcal{Q}_{a,b} +\mathcal{Q}_{b,a}\right)  = -\frac{\chi (1-\chi) k  \lambda \Delta T}{T}.
\end{equation}
The current density in the direction of the gradient is
\begin{equation}
\text{J}_{\mathcal{Q}} = \mathcal{J}_\mathcal{Q} \rho_\text{s}  = -\frac{\chi (1-\chi) k  \lambda } {T d} \nabla_{\! x}  T,
\end{equation}
and after using Eq.~(\ref{eq:Fourier}) we obtain
\begin{equation}
\label{eq:heatcond}
\kappa  = \frac{\chi (1-\chi) k  \lambda}{T d},
\end{equation}
which is the ETI thermal conductivity.

Combining the results in Eqs.~(\ref{eq:eleccond}) and (\ref{eq:heatcond})  
we arrive at a WF relation for molecular systems:
\begin{equation}
\label{eq:WFmol}
\frac{\kappa}{\sigma} =  \left(\frac{k_\text{B}}{e}\right)^2  \frac{\lambda}{k_\text{B}} = L_\text{M} T_\text{M},
\end{equation}
where $L_\text{M} = \left(k_\text{B}/ e\right)^2$ is the Lorenz number for molecular hopping conductance
and the effective temperature $T_\text{M} = \lambda/k_\text{B}$ 
is parameterized by the reorganization energy $\lambda$.
Comparing the MWF result in Eq.~(\ref{eq:WFmol}) with the WF law for metals in Eq.~(\ref{eq:WFmetal}),
we see that the explicit dependence on temperature is replaced by the temperature dependence of the reorganization energy. 
Moreover, the MWF relation is proportional to $\lambda$,
which implies that the strength of electron-phonon coupling and its manifestation through the reorganization energy dictates the relationship between ETI thermal conductivity and electrical conductivity in molecular systems.
Note that while we have drawn analogies between the functional forms of the MWF law and the standard WF law, Eq.~(\ref{eq:WFmol}) constitutes an entirely new scaling law that is valid for systems in which charge transport is dominated by inelastic electron hopping---dramatically different physics than what is used to derive the standard WF law.

\begin{figure}[]
\includegraphics[width = 8.6cm,clip]{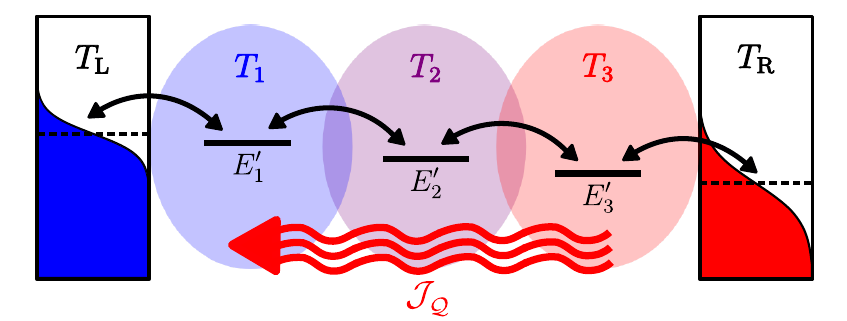}
 \caption{\label{fig:Main}
Schematic diagram of a representative molecular junction system.
The specific model we examine consists of $N$ charge transfer sites seated between two metal electrodes (rectangles).
Each site $s$ in the molecular bridge
is associated with an electronic occupation energy $E_s'$ and a set of vibrational modes that are in equilibrium with a local thermal environment at temperature $T_s$ (transparent oval).
The respective temperatures of the electrodes are $T_\text{L}$ and $T_\text{R}$.
The biased chemical potentials of the electrodes $\mu_\text{L}$ and  $\mu_\text{R}$ are shown as dashed lines
and the corresponding Fermi-Dirac distribution of each electrode is represented by a colored region.
}
\end{figure}

\begin{figure*}[t]
\includegraphics[width = 17.2cm,clip]{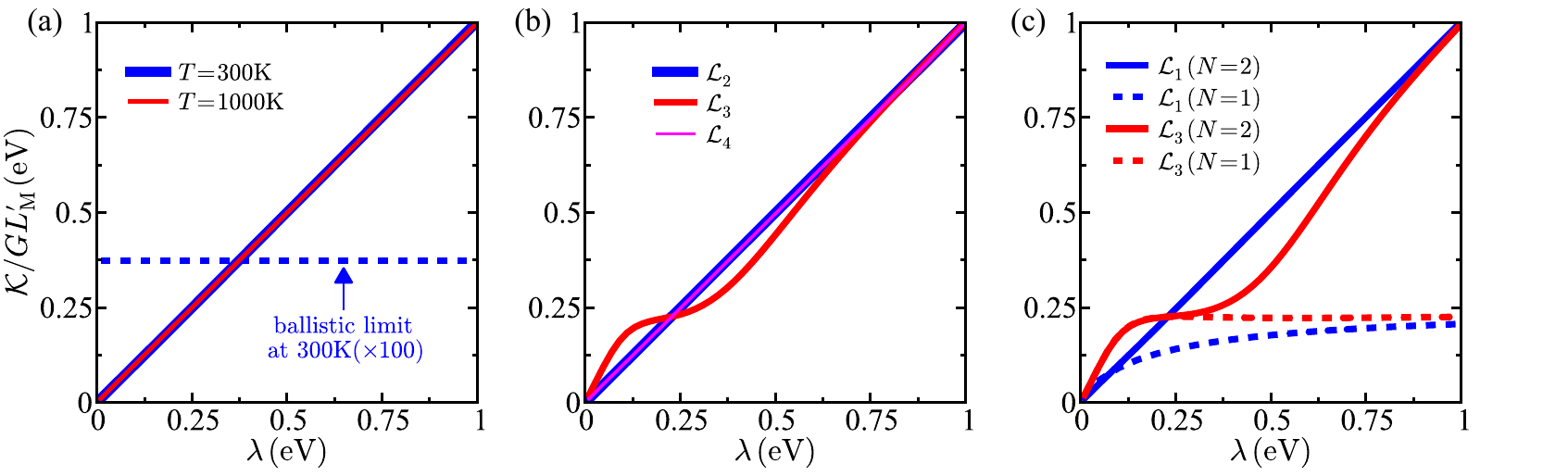}
\caption{\label{fig:WF_ER}
Calculated conductance ratio $\mathcal{K}/G L'_\text{M}$  with $L'_\text{M} = L_\text{M}/k_\text{B}$ as a function of reorganization energy $\lambda$ in a  model molecular junction.
(a)  Conductance ratios for energy landscape $\mathcal{L}_1 \equiv E'_s = 0 \, \forall  s$ (the energy levels are measured relative to $\mu$) with $N = 5$ sites. The solid red diagonal line is the hopping result for $300\,\text{K}$ and the solid blue diagonal line is the hopping result for $1000\,\text{K}$.
Note that the red line aligns over the blue line. 
The dashed blue horizontal line is the result calculated in the ballistic Landauer limit at $300\,\text{K}$ for a junction with $N = 5$ sites and landscape $\mathcal{L}_1$ (full details of this calculation are shown in Supplemental Material).
(b)
Conductance ratios in a junction with $N=5$ sites at $300\,\text{K}$ for energy landscapes $\mathcal{L}_2 \equiv E'_s = 0.25\,\text{eV} \, \forall  s$ (blue), $\mathcal{L}_3 \equiv E'_s = -0.25\,\text{eV} \, \forall s$ (red), and $\mathcal{L}_4$ (magenta) which is linear ramp of energy levels from 0.15\,\text{eV} to  -0.15\,\text{eV}  (see, for example, the level spacing in Fig.~\ref{fig:Main}). Notice that the magenta line aligns over the blue line.
(c) Conductance ratios for energy landscapes $\mathcal{L}_1$ (blue) and $\mathcal{L}_3$ (red) with $N = 2$ (solid) and $N = 1$ (dashed) sites.
In all calculations shown in this Figure, 
the electronic coupling between molecular sites is $0.01\,\text{eV}$ 
and the tunneling coupling for ET between molecule and metal is $\approx 0.07\,\text{eV}$.
} 
\end{figure*}

The robustness of the MWF law can be examined by 
calculating the WF ratio in a model molecular junction and comparing it with the general result in Eq.~(\ref{eq:WFmol}).
The specific process we consider
is electron hopping transport along a bridge of $N$ charge transfer sites seated between two metal electrodes (see Fig.~\ref{fig:Main}),
as is often invoked to examine long-range junction transport \cite{Cahen2014}.
Note that electrical conductance $G$ and thermal conductance $\mathcal{K}$
replace the corresponding conductivities as the pertinent transport properties in this model.
The respective temperatures of the electrodes are $T_\text{L} = T +\Delta T/2$ and $T_\text{R} = T -\Delta T/2$ 
where $\Delta T = T_\text{L} - T_\text{R}$ is the temperature bias across the device. 
The biased chemical potentials of the electrodes are $\mu_\text{L} = \mu - e V/2$ 
and  $\mu_\text{R} = \mu + e V/2$ where $\mu$ is the Fermi level and $V$ is an applied voltage bias across the junction.
Each site $s$ in the molecular bridge
is associated with an electronic occupation energy $E_s'$ and a local thermal environment at temperature $T_s$.
The spatial distribution of the energy levels $E'_s$ in the junction, i.e., the electronic structure of the molecular bridge, is termed an \textit{energy landscape}.

Electrons move between sites via a hopping mechanism (represented by black arrows in Fig.~\ref{fig:Main}), and
therefore the system has $N+1$ electronic states $a \in \left\{\text{M},1, \ldots, N\right\}$
where, for example, in state 1 the charge density of the hopping electron is localized on molecular site 1, 
in state 2 the charge density is localized on molecular site 2, and so forth.
The special case of $a = \text{M}$ 
corresponds to the state in which the electron
occupies an energy level in one of the metal electrodes.
Hopping between molecular sites 
and across the molecule-metal interfaces 
is controlled by 
a nuclear reorganization energy $\lambda$ that is described using the Marcus formalism \cite{craven16c, Marcus1956,Hush1961,Kuznetsov1999Electron,Nitzan2006chemical}.
The temperature profile in the molecular bridge is defined through application of a 
variant of the 
self-consistent reservoir method \cite{Bolsterli1970SCR,Bonetto2004SCR,Segal2009SCR,Tulkki2013SCR} that is modified to treat ETI heat transport \cite{note1}.

The electronic current through the junction $\mathcal{J}_\text{el}$ can be derived 
from the kinetic master equations describing the 
probability for an electron to localize on each site in the bridge or in the electronic manifold of one of the metals \cite{note1}.
The electrical conductance in the linear transport regime and under zero temperature bias ($\Delta T = 0$) is 
\begin{equation}
\quad G = \lim_{V \to 0}\frac{\mathcal{J}_\text{el}}{V} \bigg|_{\Delta T = 0},
\end{equation}
where $V$ is the voltage bias across the junction.
In the presence of a temperature bias ($\Delta T \neq 0$), ET across the thermal gradient in the molecular junction generates
a heat current $\mathcal{J}_\mathcal{Q}$ through the device (see Fig.~\ref{fig:Main}) \cite{note1},
and the corresponding thermal conductance obtained in the linear response regime under zero electric current conditions is
\begin{equation}
\mathcal{K} = \lim_{\Delta T \to 0}\frac{\mathcal{J}_\mathcal{Q}}{\Delta T} \bigg|_{\mathcal{J}_\text{el} = 0}.
\end{equation}


The conductance ratio $\mathcal{K} / G L'_\text{M}$ ($L'_\text{M} = L_\text{M}/k_\text{B}$) calculated in the molecular junction model described above is shown in Fig.~\ref{fig:WF_ER} as a function of $\lambda$.
This ratio can be compared with the MWF law which predicts that $\mathcal{K} / G L'_\text{M}  = \lambda$.
Varying the electronic landscape of the molecular bridge, as well other physical properties of the system, such as temperature, can affect the conductance ratio.
Figure~\ref{fig:WF_ER}(a) illustrates the calculated conductance ratios for energy landscape $\mathcal{L}_1 \equiv E'_s = 0 \, \forall s$ at $T = 300\,\text{K}$ (blue) and $T = 1000\,\text{K}$ (red).
At both temperatures, the numerically-calculated result almost exactly follows the MWF law. 
We have confirmed that this agreement persists for temperatures as low as our numerical methods allow sampling ($\approx T = 150\,\text{K}$) as well.
The dashed horizontal line is the conductance ratio calculated in the ballistic transport limit using the Landauer formalism \cite{note1}.
The ballistic result is independent of $\lambda$ due to the absence of electron-phonon interactions in the Landauer picture, and thus does not follow the MWF law. 
Shown in Fig.~\ref{fig:WF_ER}(b) are the results for several different energy landscapes (see the Fig.~\ref{fig:WF_ER} caption for details) calculated at $300\,\text{K}$.
Notice that in all cases the calculated results show either excellent or essentially exact agreement with the MWF law.
This illustrates the robustness of Eq.~(\ref{eq:WFmol}) for describing conductivity in molecules with varying electronic structures.
We have confirmed that this agreement is observed over a broad range of energy landscapes and temperatures (see Fig.~S1 in the Supplemental Material).
Figure~\ref{fig:WF_ER}(c) illustrates how the conductance ratio is affected by changing the number of sites in the molecular bridge.
In the case of a one site bridge ($N=1$), the MWF is not obeyed. This is expected because the MWF law is derived for molecule-to-molecule ET processes which are absent in a system with one charge transfer site. In the two-site case ($N=2$), 
the results agree with the MWF law.
This implies that for a junction with $N>1$ sites, 
the ratio between thermal and electrical conductivity is mostly dominated by intersite hopping processes in the molecular bridge.

\begin{figure}[]
\includegraphics[width = 9.6cm,clip]{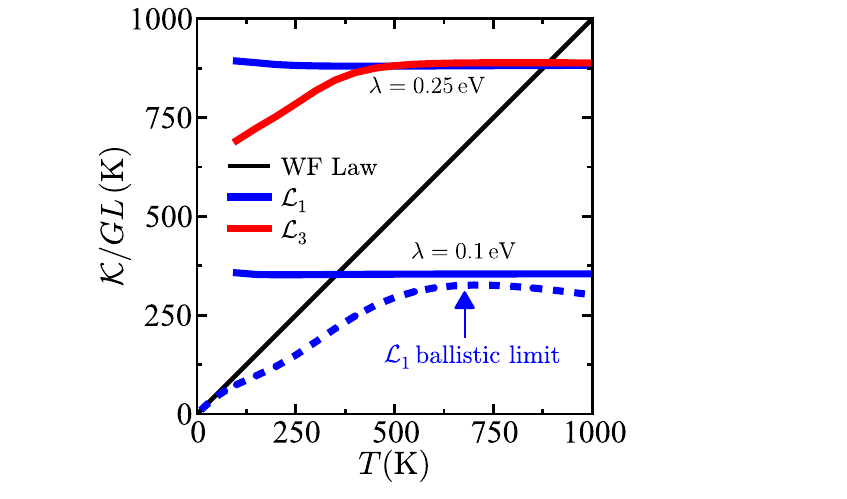}
\caption{\label{fig:WF_T}
Calculated conductance ratio $\mathcal{K}/G L$ ($L$ is the Lorenz number for metals) as a function of temperature $T$ in a model molecular junction with 
$N = 5$ sites.
The two upper solid curves are the hopping results for energy landscapes $\mathcal{L}_1$ (blue) and $\mathcal{L}_3$ (red) with reorganization energy $\lambda = 0.25\,\text{eV}$.
The lower solid curve (blue) is the result for energy landscape $\mathcal{L}_1$ and reorganization energy $\lambda = 0.1\,\text{eV}$.
In all hopping calculations, the electronic coupling between molecular sites is $0.01\,\text{eV}$ and the tunneling coupling for ET between molecule and metal is $\approx 0.07\,\text{eV}$.
The dashed blue curve is the result for energy landscape $\mathcal{L}_1$ in the ballistic Landauer limit with molecule-molecule electronic coupling $0.1\,\text{eV}$
and molecule-metal coupling $\approx 0.7\,\text{eV}$. 
The black diagonal line is the result predicted by the WF law.
}
\end{figure}

A comparison between the temperature-dependence of the conductance ratio $\mathcal{K} / G L$ in the molecular junction model and the ratio predicted by the traditional WF law ($\mathcal{K} / G L = T$) is shown in Fig.~\ref{fig:WF_T}.
The solid curves illustrate the numerical results for different energy landscapes and reorganization energies. 
The dashed curve is the result of a ballistic Landauer calculation.
Three observations are noteworthy:
(a) the calculated results deviate significantly from the WF law,
(b) the ballistic result follows the WF law only at very low temperatures, which agrees with previous results \cite{Beneti2012},
and
(c) the hopping results are either approximately temperature-independent (assuming temperature-independent reorganization energies)
over all temperature ranges (solid blue curves) or weakly temperature-dependent in the low-temperature regime while becoming temperature-independent at higher temperatures (solid red curve). Note that for most energy landscapes we have examined, the conductance ratio associated with hopping transport is approximately temperature-independent (see Fig.~S2 in the Supplemental Material), in agreement with the MWF law.

We have derived a WF law for molecules and confirmed its validity using numerical simulations of a 
paradigmatic molecular nanostructure.
It is significant that the MWF law is able to accurately describe conductivity relations in hybrid molecule-metal systems.
We therefore expect that it will describe the relationship between electrical conductivity and ETI thermal conductivity
in both purely molecular systems and in systems where electron transfer takes place between heterogeneous structures
with molecular components.
Experimental examination of the theoretical results presented here could be realized using similar setups to those that are currently
applied to probe the thermal, electronic, and thermoelectric properties of molecular junctions \cite{Reddy2007,Tan2011,Lee2013,Kim2014,Venkataraman2015,Garner2018, Reddy2019nature}.
The MWF law opens a new way to analyze conductance in molecular systems, 
and could lead to advancements in the design of thermoelectric and photovoltaic devices, computing architectures, and molecular electronics.


The research of AN is supported by the Israel-U.S. Binational Science Foundation, 
the German Research Foundation (DFG TH 820/11-1), 
the U.S. National Science Foundation (Grant No. CHE1665291),
and the University of Pennsylvania.

\bibliography{j,c5,electron-transfer,networks,craven,osc-bar,pensphere,nonequilibrium}
\end{document}